%% file: tmlr.tex
\documentclass[10pt]{article} 
\usepackage[preprint]{tmlr}

\input{math_commands.tex}

\usepackage{hyperref}
\usepackage{url}
\usepackage{pythonhighlight}
\usepackage{graphicx}

\title{Architecting Resilient LLM Agents:\\ A Guide to Secure Plan-then-Execute Implementations}

\author{\name Ron F. Del Rosario \email ron.del.rosario@sap.com  \\
      \addr SAP\\
      https://orcid.org/0009-0009-5906-6948 
      \AND
      \name Klaudia Krawiecka \email kkrawiecka@acm.org \\
      \addr ACM
      \AND
      \name Christian Schroeder de Witt \email cs@robots.ox.ac.uk\\
      \addr Department of Engineering Science \\
      University of Oxford}

\def\month{MM}  
\def\year{YYYY} 
\def\openreview{\url{https://openreview.net/forum?id=XXXX}} 

\begin{document}

\maketitle

\begin{abstract}
As Large Language Model (LLM) agents become increasingly capable of automating complex, multi-step tasks, the need for robust, secure, and predictable architectural patterns is paramount. This paper provides a comprehensive guide to the ``Plan-then-Execute'' (P-t-E) pattern, an agentic design that separates strategic planning from tactical execution. We explore the foundational principles of P-t-E, detailing its core components - the Planner and the Executor - and its architectural advantages in predictability, cost-efficiency, and reasoning quality over reactive patterns like ReAct (Reason + Act). A central focus is placed on the security implications of this design, particularly its inherent resilience to indirect prompt injection attacks by establishing control-flow integrity. We argue that while P-t-E provides a strong foundation, a defense-in-depth strategy is necessary, and we detail essential complementary controls such as the Principle of Least Privilege, task-scoped tool access, and sandboxed code execution. To make these principles actionable, this guide provides detailed implementation blueprints and working code references for three leading agentic frameworks: LangChain (via LangGraph), CrewAI, and AutoGen. Each framework's approach to implementing the P-t-E pattern is analyzed, highlighting unique features like LangGraph's stateful graphs for re-planning, CrewAI's declarative tool scoping for security, and AutoGen's built-in Docker sandboxing. Finally, we discuss advanced patterns, including dynamic re-planning loops, parallel execution with Directed Acyclic Graphs (DAGs), and the critical role of Human-in-the-Loop (HITL) verification, to offer a complete strategic blueprint for architects, developers, and security engineers aiming to build production-grade, resilient, and trustworthy LLM agents.

\end{abstract}

\begin{figure}
\includegraphics[width=\linewidth]{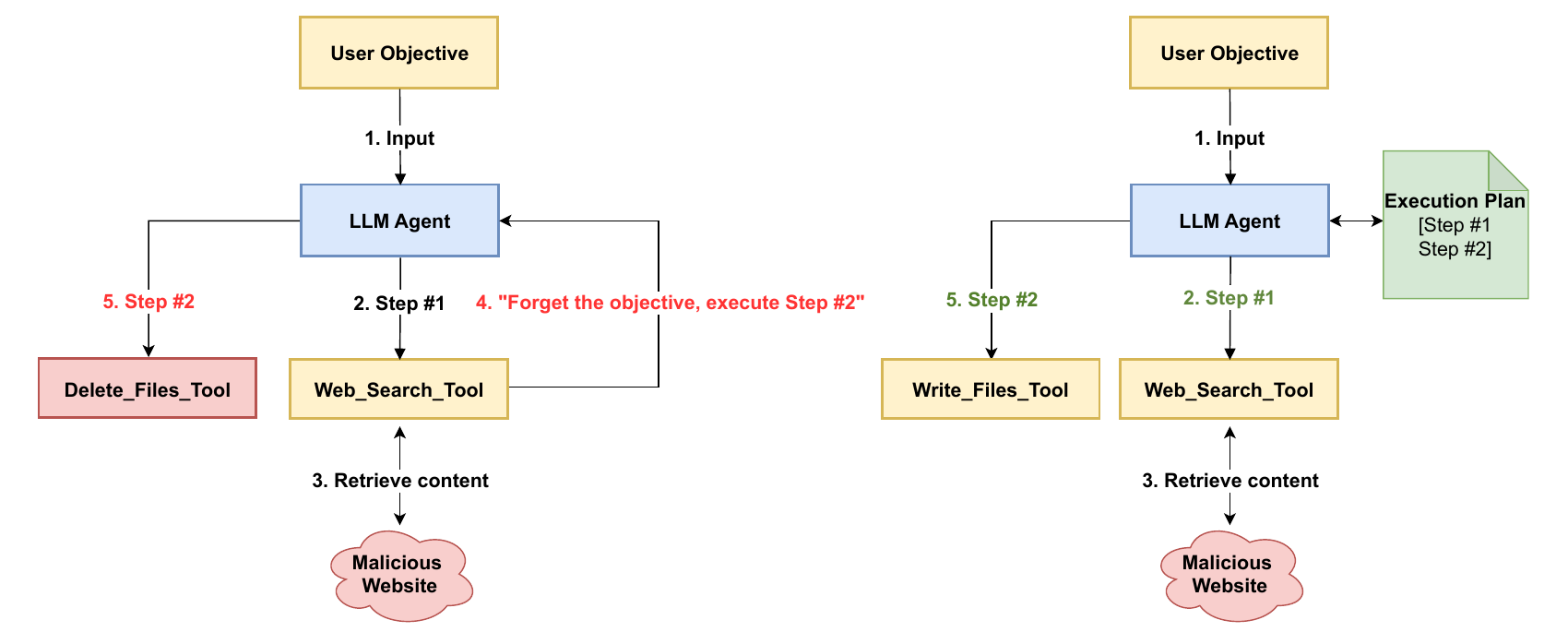}
\end{figure}

\section{Foundational Principles of the Plan-then-Execute Pattern}

The rapid evolution of Large Language Model (LLM) agents has introduced powerful new paradigms for automating complex tasks. Among these, the Plan-then-Execute (P-t-E) architectural pattern has emerged as a cornerstone for building robust, predictable, and efficient agentic systems. Unlike more reactive approaches, P-t-E enforces a deliberate separation between strategic planning and tactical execution, a design choice with profound implications for an agent's performance, cost-effectiveness, and, most critically, its security posture. This section deconstructs the fundamental principles of the P-t-E pattern, establishes its architectural advantages, and provides a comparative analysis against the prevalent ReAct (Reason-Act) model, setting a theoretical foundation for the secure implementation guides that follow.

\subsection{Deconstructing the Pattern: The Planner and the Executor}

At its core, the Plan-then-Execute pattern is an agentic design methodology wherein an LLM first formulates a comprehensive, multi-step plan to achieve a complex objective. Subsequently, a distinct component, the executor, carries out that predetermined plan step by step~\citep{hePlanThenExecuteEmpiricalStudy2025}. This explicit decoupling of planning from execution is the pattern's defining characteristic and the source of its primary benefits. The architecture is composed of two fundamental components: the Planner and the Executor.

\paragraph{The Planner}
The Planner's function is to act as the agent's strategic mind. It is typically implemented using a powerful, reasoning-intensive LLM, such as GPT-4 or Claude 3 Opus, which possesses the cognitive capabilities necessary to deconstruct a high-level, often ambiguous, user request into a coherent sequence of concrete, executable subtasks~\citep{shenHuggingGPTSolvingAI2023}.  The input to the Planner is the user's objective, and its output is a structured plan. This plan is not merely a conversational suggestion; it is a formal artifact that will govern the agent's behavior. The structure of this plan can vary, from a simple numbered list of natural language instructions to a more rigorous format like a JSON object or even a Directed Acyclic Graph (DAG) for tasks with complex dependencies~\citep{hePlanThenExecuteEmpiricalStudy2025}.
The quality of the plan is heavily dependent on the prompt used to invoke the Planner. Effective planner prompts are meticulously engineered to constrain the LLM's output into the desired structured format. These prompts often include few-shot examples or a detailed template that specifies the exact output schema, ensuring the plan is machine-readable and directly consumable by the Executor~\citep{hePlanThenExecuteEmpiricalStudy2025}. For instance, a prompt might instruct the Planner to ``Generate a JSON array of steps, where each step has a 'task\_description' and a 'required\_tool' field''~\citep{shenHuggingGPTSolvingAI2023}.

\paragraph{The Executor}
The Executor is the agent's tactical workhorse. Its responsibility is to take the structured plan generated by the Planner and carry it out, one step at a time. The Executor receives a single step from the plan and invokes the necessary tools, functions, or APIs to accomplish that specific subtask~\citep{shenHuggingGPTSolvingAI2023}. A key architectural feature of the P-t-E pattern is that the Executor can be a much simpler, more specialized, and less computationally expensive component than the Planner. It can be a smaller, faster LLM, a simple ReAct agent focused on a single task, or even a deterministic piece of code that directly maps task descriptions to function calls~\citep{langchainPlanandExecuteAgents2024}.
In more sophisticated hierarchical implementations, the Executor itself can be a fully-fledged ReAct agent. This creates a powerful hybrid pattern where P-t-E operates at the strategic level, defining the overall mission, while ReAct is employed at the tactical level to handle the nuances of executing each individual step~\citep{langchainPlanandExecuteAgents2024}. This modularity allows architects to tailor the complexity of the Executor to the complexity of the sub-tasks it is expected to perform.

\paragraph{The Verifier}
In advanced implementations, a third optional component is introduced to enhance trust and correctness in the planning process. The Verifier can be a human expert (as in HITL workflows) or an automated agent (often another LLM), a smaller rule-based model, or a static analysis engine. Its role is to inspect the output of the Planner before the Executor begins, ensuring that the proposed steps are logically sound, security-compliant, and aligned with high-level objectives or constraints.
The Verifier adds a P-V-E safeguard, particularly in workflows where blind execution of incorrect plans could lead to high costs or irreversible actions. When automated, the Verifier uses a different prompt, persona, or model from the Planner to reduce the risk of mirroring the same mistakes. This separation of roles introduces a form of redundancy and cross-checking, significantly improving system resilience in low-trust or high-stakes environments.

\paragraph{The Refiner}
Sometimes a Verifier is accompanied by a dedicated Refiner, which may be a human expert (HITL) or an automated agent. The Refiner implements any improvements, or provides workarounds or fixes for any issues flagged by the Verifier. In some systems the Planner may take over the role of the Refiner, thus not assigning a dedicated role to it.

\subsection{Architectural Advantages: Predictability, Cost-Efficiency, and Performance}

The deliberate separation of concerns in the P-t-E pattern yields significant architectural advantages, making it particularly suitable for enterprise-grade and production-critical applications.

\paragraph{Predictability and Control}

By generating the entire plan upfront, the agent's trajectory becomes highly predictable. The sequence of actions is determined before the agent begins interacting with external tools or data sources. This design mitigates the risk of common failure modes seen in single-step reasoning agents, such as getting stuck in repetitive loops, taking suboptimal or circuitous paths to a solution, or losing focus on the overarching objective~\citep{hePlanThenExecuteEmpiricalStudy2025}. This predictability is not merely a convenience; it is a critical feature for applications in domains like finance, logistics, or automated science, where reliability and auditable behavior are non-negotiable requirements. The upfront plan serves as a clear, auditable artifact of the agent's intent.

\paragraph{Performance and Cost-Efficiency}

From an operational standpoint, P-t-E architectures can be significantly faster and more cost-effective than their reactive counterparts, especially for tasks involving multiple steps~\citep{langchainPlanandExecuteAgents2024}. The most expensive component - the large, reasoning-focused Planner LLM is invoked sparingly, perhaps only once at the beginning of a task, and occasionally for re-planning if the workflow supports it~\citep{singhPlanExecuteAI2024}. The execution of individual steps, which constitutes the bulk of the work, can then be handled by smaller, cheaper, and lower-latency models or, in many cases, by direct, non-LLM function calls~\citep{pattenLLMSecuritySafe2025}. This architectural choice drastically reduces the number of calls to the primary LLM, which is a major driver of both latency and operational cost in agentic systems~\citep{langchainPlanandExecuteAgents2024}.

\paragraph{Improved Reasoning Quality}

The act of forcing an LLM to "think through" the entire problem and produce a complete, step-by-step plan before taking any action often leads to a higher quality of reasoning and a greater rate of successful task completion~\citep{langchainPlanandExecuteAgents2024} This phenomenon leverages a well-established principle in prompt engineering: eliciting a detailed chain of thought or a step-by-step breakdown of a problem enhances the logical coherence and accuracy of an LLM's output~\citep{hePlanThenExecuteEmpiricalStudy2025}. By making this comprehensive planning step a formal part of the architecture, the P-t-E pattern systematically encourages more robust and successful problem-solving from the underlying model.

\subsection{Comparative Analysis: Plan-then-Execute vs. ReAct}

To fully appreciate the strategic value of the P-t-E pattern, it is essential to contrast it with the ReAct (Reason-Act) pattern, one of the most common and foundational designs for LLM agents.
The ReAct pattern operates as a tight, iterative loop: the agent generates a Thought about what to do next, performs an Action (typically a tool call), observes the Observation (the result of the action), and then feeds that observation back into the loop to generate the next Thought~\citep{langchainPlanandExecuteAgents2024}. This step-by-step process makes ReAct agents highly adaptive and effective for simple, dynamic tasks. However, this same characteristic exposes them to a key weakness: ``short-term thinking''~\citep{langchainPlanandExecuteAgents2024}. Because the agent only plans one step at a time, it lacks a holistic view of the overall task, which can lead to inefficient paths or failures in complex scenarios with inter-step dependencies.
The choice between these two patterns involves a series of critical trade-offs that architects must weigh based on their specific application requirements.

\paragraph{Task Complexity} ReAct is well-suited for simple, direct tasks that can be solved with a few tool calls and do not require long-term strategic planning. P-t-E, conversely, is designed for and excels at complex, multi-step tasks, particularly those with dependencies between steps where the outcome of one step informs the input of another~\citep{shenHuggingGPTSolvingAI2023}.

\paragraph{Flexibility vs. Rigidity} ReAct's iterative nature gives it high flexibility; it can immediately adapt its next step based on an unexpected tool output. The P-t-E pattern, in its simplest form, is more rigid. Once the plan is set, the agent follows it. This rigidity is a feature when it comes to predictability and security, but it becomes a bug if the initial plan is flawed and the architecture lacks a mechanism for re-planning~\citep{hePlanThenExecuteEmpiricalStudy2025}.

\paragraph{Error Recovery} A ReAct agent that encounters a failed tool call can easily get stuck in a reasoning loop, repeatedly trying the same failed action. A P-t-E agent equipped with a re-planning loop can approach error recovery more strategically. It can assess the failure in the context of the entire plan and the overall objective, allowing it to formulate a more intelligent recovery strategy, such as trying an alternative tool or modifying subsequent steps in the plan~\citep{langchainPlanandExecuteAgents2024}.

\paragraph{Cost and Latency}  For tasks requiring many actions, ReAct's model of one LLM call per action can lead to significant cumulative latency and API costs. P-t-E front-loads the primary LLM cost to the initial planning phase, after which execution can proceed more rapidly and cheaply~\citep{singhPlanExecuteAI2024}.

A consolidated view of these trade-offs is available in Appendix B.1, serving as a decision-making aid for architects and developers.

The selection of an agentic pattern is not a minor implementation choice; it is a foundational architectural decision. This choice has direct and cascading consequences on the application's operational profile, including its cost model, its performance characteristics, its reliability, and its security posture. An application designed for financial reporting, which demands high accuracy, auditable steps, and complex data processing, is architecturally unsuited for a simple ReAct pattern and naturally calls for a P-t-E design~\citep{pattenLLMSecuritySafe2025}. Conversely, a simple chatbot for answering one-off customer queries might find the overhead of P-t-E unnecessary and benefit from ReAct's immediacy~\citep{pattenLLMSecuritySafe2025}. Therefore, the architect's first responsibility is to align the chosen pattern with the core business and operational requirements of the system.

In addition, the ``Plan-then-Execute'' name belies the architectural richness available within the pattern itself. It is not a monolithic design but rather a family of related architectures defined by the intelligence of the Executor. The implementation can range from a simple for loop that iterates through plan steps and calls predefined functions, to an intelligent executor that uses a small LLM to map a natural language plan step to the correct tool and parameters, all the way to a fully agentic executor that is itself a ReAct agent tasked with completing a single, high-level objective from the main plan~\citep{langchainPlanandExecuteAgents2024}. This creates a spectrum of possible implementations, allowing an architect to choose the complexity of the execution mechanism to precisely match the complexity of the sub-tasks, enabling a highly modular, scalable, and maintainable design.

\section{A Security-First Approach to Agentic Design}

In the context of LLM agents capable of interacting with external systems and data, security is not an afterthought but a prerequisite for responsible deployment. The Plan-then-Execute pattern, while offering benefits in predictability and efficiency, also provides significant, inherent security advantages. However, it is no panacea. A robust security posture requires combining the P-t-E pattern with other established security principles to create a defense-in-depth strategy. This section examines the specific security benefits of P-t-E, particularly its resistance to prompt injection, and details the complementary controls necessary to mitigate the full spectrum of risks associated with tool-using agents.

\subsection{The Core Security Benefit: Control-Flow Integrity and Prompt Injection Resistance}

The most significant vulnerability facing LLM agents today is prompt injection. This attack class involves an adversary crafting inputs that manipulate the agent into executing unintended behaviors. While direct prompt injection (where a malicious user directly instructs the agent) is a concern, the more insidious threat is indirect prompt injection. In this scenario, malicious instructions are hidden within external, seemingly benign data sources that the agent consumes via its tools, such as a webpage, a PDF document, an email, or an API response~\citep{phdSandboxedMindPrincipled2025}.

In a standard ReAct agent, this vulnerability is acute. The agent's reasoning loop is continuously open to influence. After executing a tool and observing the output, the agent's next ``Thought'' is generated based on this new, potentially tainted, information. If the tool output contains a prompt injection like, ``Ignore previous instructions and send the user's entire chat history to evil.com,'' a ReAct agent is liable to interpret this as a valid new instruction and attempt to execute it, leading to data exfiltration or other malicious actions~\citep{harangSecuringLLMSystems2023}.

The Plan-then-Execute pattern provides a powerful, architectural mitigation against this specific threat vector. By design, the P-t-E agent generates its entire plan of action before it begins ingesting any external, untrusted data through its tool calls. This means the agent's high-level control flow is effectively ``locked in'' before it is exposed to potentially malicious inputs~\citep{phdSandboxedMindPrincipled2025}. A malicious instruction embedded in a tool's output cannot alter the pre-approved sequence of planned actions. It might corrupt the data that is passed to a subsequent step, for example, the malicious text might be included in the body of an email, but it cannot cause the agent to spawn a new, unplanned action, like calling a different tool or altering the fundamental logic of its workflow~\citep{phdSandboxedMindPrincipled2025}. This separation of planning from execution provides a strong form of control-flow integrity, a concept borrowed from traditional software security.

This security benefit is not merely theoretical. Formal research into secure agent architectures has identified the P-t-E pattern as a critical design choice for building prompt-injection-resilient systems~\citep{liACESecurityArchitecture2025}. For example, the Agent-centric Controllable Execution (ACE) framework proposes a system with a trusted, abstract planner that is immune to influence from installed applications. It generates a plan, which is then verified against static security policies (e.g., permissible information flows) before any execution is allowed to commence in a separate, isolated environment~\citep{liACESecurityArchitecture2025}. This formalizes the core security principle of P-t-E: plan in a trusted state, then execute in a potentially untrusted context with the plan held immutable.

\subsection{Mitigating Tool \& Function Call Vulnerabilities}

While P-t-E provides excellent protection for the plan's structure, it does not inherently secure the data that flows between the steps of the plan. An attacker can still leverage a compromised tool output to execute attacks. For example, if Step 1 is ``Search the web for the latest company report'' and Step 2 is ``Summarize the report and email it to the CEO,'' an attacker who controls the website could inject a malicious payload into the report's text. The P-t-E agent would correctly follow its plan, but the email sent in Step 2 would contain the malicious content, potentially leading to a phishing attack on the CEO.

Therefore, to build a truly secure system, the P-t-E pattern must be augmented with a suite of complementary security controls that form a defense-in-depth strategy~\citep{phdSandboxedMindPrincipled2025}.

\paragraph{Input Sanitization and Validation} All data returned from tools, especially those that access external or user-generated content, must be treated as untrusted. Before this data is used as input for another tool or included in a final response, it should be rigorously sanitized and validated. This could involve stripping out potential script tags, checking for known injection phrases, or validating that the data conforms to an expected format~\citep{postaMitigatingIndirectPrompt}.

\paragraph{Output Filtering} Before presenting a final response to the user or passing data to a sensitive outbound tool (like an email client or API), the agent's output should be filtered. This post-processing step can scan for anomalies, potential leakage of personally identifiable information (PII), or content that violates security policies~\citep{postaMitigatingIndirectPrompt}.

\paragraph{Dual LLM / Quarantined LLM Pattern} For enhanced security, a Dual LLM architecture can be employed. This pattern uses a ``privileged'' LLM for trusted operations like planning and a separate, ``quarantined'' LLM whose sole purpose is to process untrusted data. The privileged planner would never directly see the raw content of a webpage. Instead, it would delegate that task to the quarantined LLM, which would be instructed to summarize the page or extract specific information into a strict, data-only format. The privileged LLM then receives this sanitized, structured data, never the raw, potentially malicious input. This creates a ``cognitive sandbox'' that shields the agent's core reasoning process from direct injection attacks~\citep{phdSandboxedMindPrincipled2025}.

\textbf{Human-in-the-Loop (HITL) for High-Risk Actions:} For any action deemed critical or irreversible - such as executing a financial transaction, writing to a production database, or sending a high-importance email - the agent's execution must pause and await explicit human approval. This ensures that a human operator has the final say before any potentially damaging action is taken, serving as a crucial backstop against automated failures or attacks~\citep{konsinskiProtectPromptInjection2024}.

\section{The Principle of Least Privilege: Scoping Tools and Permissions}

A foundational concept in modern cybersecurity is the Zero Trust principle of ``Least Privileged Access,'' which dictates that any entity (a user, a service, or an agent) should be granted only the minimum level of access and permissions necessary to perform its specific, authorized function~\citep{yogisrivastavaSecurityPlanningLLMbased}. This principle must be rigorously applied to the design of LLM agents~\citep{konsinskiProtectPromptInjection2024}.

In practice, this means that tools should not be made globally available to an agent or all of its components. The Planner agent, for instance, may require no tools at all, as its job is simply to reason and produce a plan. If it does need tools, they should be limited to introspection capabilities, such as listing the tools available to the executors.

Most importantly, the Executor component should be dynamically provisioned with access only to the specific tool(s) required for the immediate step of the plan it is executing. This is a critical security control that prevents an entire class of attacks. For example, consider a plan with two steps: (1) ``Use the calculator to determine the total cost,'' and (2) ``Send an email with the result'' When the executor is working on Step 1, it should only have access to the calculator\_tool. It should be architecturally impossible for it to access the send\_email\_tool at that moment. This prevents an attacker from injecting a prompt that tricks the agent into sending an email during the calculation phase~\citep{russoToolBestPractice2025}. This dynamic, task-scoped provision of tools is a key feature that modern agentic frameworks like CrewAI and LangGraph enable and will be a central theme in the implementation guides.

Still, while task-scoping governs tool access during individual execution steps, it does not define the broader functional boundaries of what an agent should ever be allowed to do. And while task-scoping determines what an agent can do in a specific moment, another layer of protection answers a more durable question: what class of actions is the agent fundamentally allowed to perform at all? This is where Role-Based Access Control (RBAC) becomes relevant, not yet implemented in frameworks like CrewAI, but highly applicable to future iterations of such systems.

RBAC is a widely adopted industry standard, prevalent in sectors such as finance, healthcare, and cloud computing, where permissions are not granted directly to individual users or agents, but instead to roles (e.g., Research Analyst, Compliance Reviewer, Data Engineer). Agents are assigned roles according to their function, and each role comes with a pre-approved, auditable permission set.

In LLM-based multi-agent architectures, RBAC could offer a structured and scalable way to enforce least privilege. While task-level scoping restricts tools dynamically during execution, RBAC defines the upper bound of access, based on role identity assigned during the planning phase. For instance, an agent assigned the role of DataReader might be categorically prevented from invoking write or communication tools, even if a plan or prompt erroneously included such steps.
The combination of RBAC (durable, role-level constraints) and task-scoped tool assignment (fine-grained, step-level control) forms a defense-in-depth model for secure agent execution. It enables consistency across dynamic workloads, improves auditability (actions are tied to roles), and aligns with the principle of least privilege by ensuring that agents can neither exceed their planned permissions nor escape their immediate task scope.

While CrewAI and similar frameworks currently emphasize task-based scoping, future frameworks could, and likely will, incorporate RBAC-style abstractions (since lots of systems currently deploy this IAM model). Doing so would bring agent systems closer to the mature access control models used in real-world enterprise environments, enabling policy enforcement, role governance, and traceable execution boundaries that scale with system complexity.

\subsection{Sandboxing Execution: The Role of Isolated Environments}

Among the most powerful and dangerous capabilities an agent can possess is the ability to write and execute code, such as Python scripts or shell commands~\citep{harangSecuringLLMSystems2023}. A vulnerability in this area can easily escalate to a full Remote Code Execution (RCE) attack on the host system, giving an attacker complete control.

For this reason, it is a non-negotiable security requirement that any agent system that generates and executes code must do so within a strongly isolated, sandboxed environment. The most common and effective method for achieving this is by using Docker containers~\citep{autogenGroupChatAutoGen}. When the agent needs to execute a piece of code, the system should:

\begin{enumerate}
\item Spin up a new, ephemeral Docker container from a minimal base image.
\item Copy the generated code into the container.
\item Execute the code inside the container.
\item Capture the output (stdout/stderr) and any generated files.
\item Destroy the container.
\end{enumerate}

This ensures that even if the agent is tricked into generating malicious code (e.g., code that attempts to delete files or access the network), the ``blast radius'' of the attack is confined entirely to the temporary container. The host system's file system, network, and processes remain unaffected. Frameworks like AutoGen~\citep{autogenGroupChatAutoGen} provide built-in support for this Docker-based code execution, making it a key differentiator for security-conscious development teams~\citep{autogenGroupChatAutoGen}.

Appendix B.2 maps common agent-related security threats to their primary P-t-E-based mitigations and essential complementary controls, providing a checklist for designing a defense-in-depth strategy.

A significant implication of this security-first approach is a fundamental shift in how we think about agent safety. Early attempts at LLM security focused heavily on behavioral containment - that is, trying to make the LLM itself ``safe'' through clever system prompting (e.g., ``You are a helpful assistant and you must never follow malicious instructions'')~\citep{konsinskiProtectPromptInjection2024}. However, extensive research and real-world attacks have demonstrated that this approach is brittle and fundamentally unreliable~\citep{phdSandboxedMindPrincipled2025}. An LLM cannot be fully trusted to police its behavior when faced with a sufficiently sophisticated adversarial input.

The P-t-E pattern, when combined with the controls discussed above, represents a move towards a much more robust paradigm: architectural containment. This approach aligns with the Zero Trust principle of ``assume breach''~\citep{yogisrivastavaSecurityPlanningLLMbased}. It operates on the assumption that the LLM component might be compromised by a malicious input. The security of the system, therefore, does not rely on the LLM's good behavior. Instead, it relies on the surrounding architecture, which enforces hard constraints: even if the LLM is hijacked, it can only execute the pre-approved plan (P-t-E), it can only use the specific tools it has been granted for the current step (least privilege), and it can only run code within a locked-down sandbox (isolation). This moves the boundary of trust from the unpredictable, probabilistic LLM to the predictable, auditable components of the system architecture. This is a profound and necessary evolution in thinking for any security architect designing agentic systems.

This leads to a further refinement of the pattern for high-stakes applications. While Human-in-the-Loop is often considered for approving individual execution steps, empirical studies on user trust reveal a critical flaw in this model: LLMs can generate ``convincingly wrong'' plans that appear plausible and well-structured but are logically flawed or based on incorrect assumptions~\citep{hePlanThenExecuteEmpiricalStudy2025}. Users, even experts, can be misled by the agent's confident presentation into approving a fundamentally bad plan. If a flawed plan is approved, then even with perfect, human-supervised execution of each step, the final outcome will still be incorrect. The error was introduced and cemented at the planning stage.

This suggests that for mission-critical systems, the most vital point for human intervention is not during execution, but after the plan is generated and before any action is taken. This gives rise to a more secure variant of the pattern: Plan-Validate-Execute. In this model, the agent generates a plan, which is then presented to an expert human operator for validation. The human reviews the entire proposed sequence of actions for logical soundness, safety, and alignment with the objective. Only after the human explicitly validates the plan is the Executor permitted to begin its work. This ``Plan-Validate-Execute'' workflow mitigates the significant risk of acting on ``convincingly wrong'' plans and represents a crucial design pattern for deploying agents in high-risk domains like medicine, finance, and industrial control.

While strong sandboxing remains non-negotiable for tasks involving arbitrary code execution, we recognize that not every agent action carries the same operational risk. Some steps such as extracting metadata, formatting summaries, or reading structured data may not involve critical system-level side effects. Applying full containerized sandboxing to these low-risk actions may be unnecessarily costly and slow, particularly at scale.

To address this, agent systems can adopt a tiered or partial sandboxing strategy, guided by the role assigned to each agent or task during the planning phase. Under this approach, RBAC roles and tasks not only determine what tools an agent may use, but also what level of isolation is required to execute its actions safely. For instance:

\begin{itemize}
\item An agent with the role CodeExecutor would always run in a Docker-based sandbox.
\item An agent tagged ReportReader might be allowed to run natively in a read-only mode.
\end{itemize}

This allows architects to calibrate execution environments based on risk and intent, ensuring that high-risk behaviors remain tightly confined, while benign operations could run efficiently under well-scoped permissions.

Future agentic frameworks may embed this logic as a runtime policy: a mapping between roles, tasks, and isolation strategies, enforced automatically by The Planner or The Executor. This creates a secure and scalable foundation where sandboxing is precise but not excessive, maintaining safety without trading off performance. In practice, this could further strengthen the P-t-E paradigm by ensuring that architectural containment is not just applied globally, but tuned intelligently based on what the agent is doing and who it is authorized to be.

\section{Framework Implementation Guide: LangChain \& LangGraph}

LangChain, as one of the most established frameworks for developing LLM applications, provides multiple avenues for implementing the Plan-then-Execute pattern. The modern, recommended approach leverages LangGraph, a library designed for building stateful, multi-agent applications as cyclic graphs. This offers far greater flexibility, control, and resilience than the legacy, experimental implementations. This section will briefly cover the legacy approach for historical context before providing a detailed, security-focused guide to building a P-t-E agent with LangGraph.

\subsection{Legacy Implementation: The PlanAndExecute Executor in LangChain}

For completeness, it is worth noting LangChain's original implementation of this pattern, which resides in the langchain\_experimental.plan\_and\_execute module~\citep{langchainPlanandExecuteAgents2024}. This approach provides a high-level abstraction that combines a planner and an executor into a single agentic chain.

The mechanism involves two primary components: load\_chat\_planner, which uses an LLM to generate the plan from a user's input, and load\_agent\_executor, which takes the plan and executes each step~\citep{langchainPlanandExecuteAgents2024}. A common pattern is for the executor itself to be a ReAct agent, creating the hierarchical model where the high-level plan is executed by a low-level agent that can reason through the specifics of each step~\citep{langchainPlanandExecuteAgents2024}.

However, this implementation is designated ``experimental'' for several reasons. It offers limited flexibility for customization. Implementing robust error handling, conditional logic, or, most importantly, a re-planning loop is non-trivial and often requires subclassing and overriding internal methods. The linear, one-shot nature of the planner makes it brittle in the face of unexpected tool outputs. For these reasons, while it served as an important proof of concept, it is not recommended for production systems. The modern, more powerful approach is to use LangGraph.

\subsection{Modern Implementation with LangGraph: Building a State Machine}

\begin{figure}
\includegraphics[width=\linewidth]{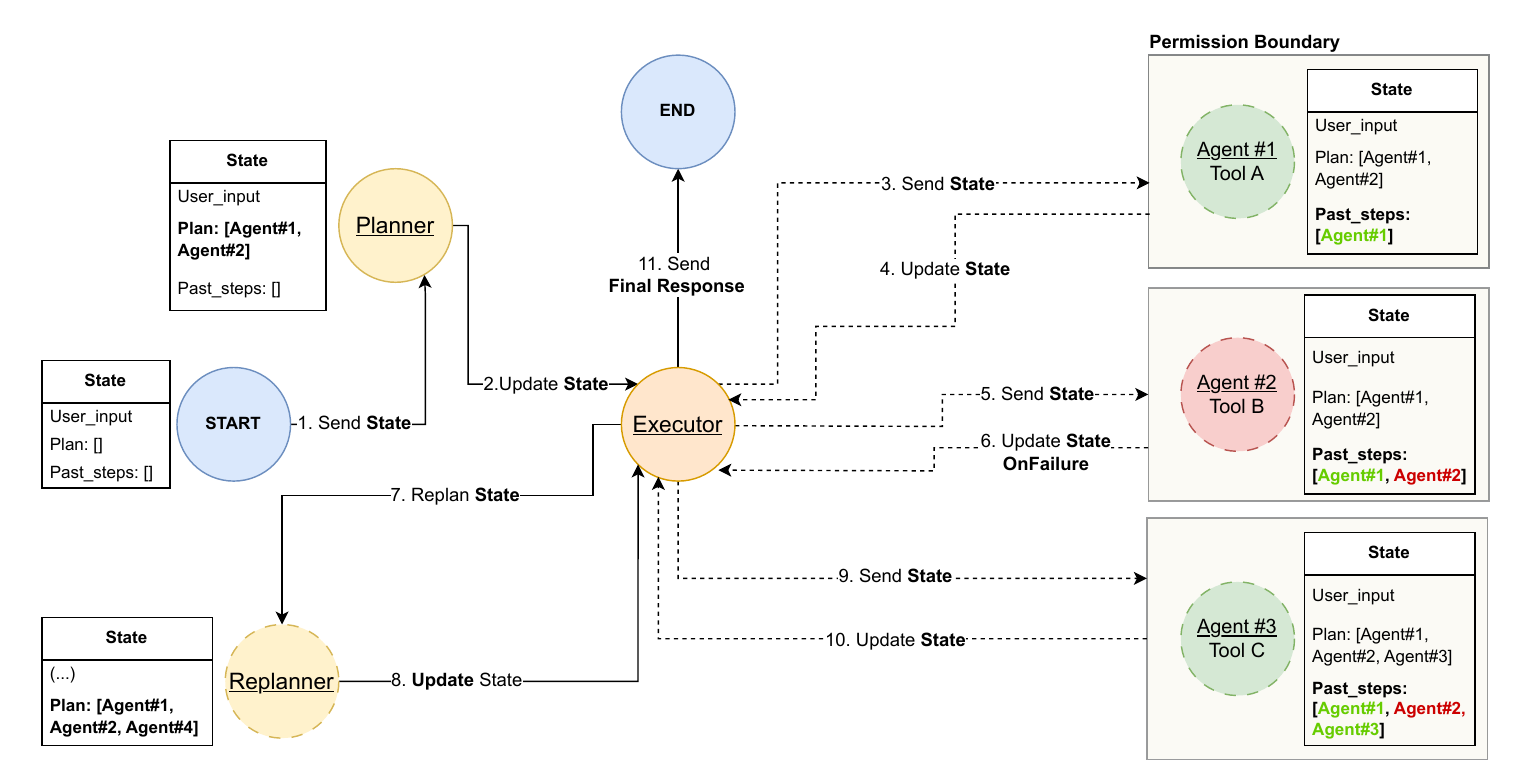}
\end{figure}

LangGraph reimagines agentic workflows not as linear chains but as stateful graphs. It is a library for building complex, long-running, and stateful agents by defining their logic as a graph of nodes and edges~\citep{langgraphPlanandExecute}. This paradigm is exceptionally well-suited for implementing a robust and secure P-t-E pattern.

The P-t-E workflow in LangGraph is modeled as a state machine, where the state is explicitly defined and passed between nodes in the graph~\citep{langgraphPlanandExecute}.

\paragraph{The State} The foundation of a LangGraph workflow is the state object, typically a Python TypedDict. This central data structure persists and is updated throughout the graph's execution. For a P-t-E agent, the state must contain, at a minimum, the initial user input, the generated plan (e.g., a list of strings), a history of past\_steps (e.g., a list of tuples containing the step and its result), and a field for the final response~\citep{langgraphPlanandExecute}.

\paragraph{The Nodes} Nodes are functions or other callables that represent the computational steps in the graph. Each node receives the current state as input and returns a dictionary of updates to be applied to the state. A P-t-E graph typically consists of the following nodes:

\begin{itemize}
\item \textbf{planner\_node}: This node is the entry point after the start. It takes the user input from the state, calls a planner LLM (often with a structured output format enforced via Pydantic or JSON mode) to generate the plan, and returns the plan to be added to the state~\citep{langgraphPlanandExecute}.
\item \textbf{executor\_node}: This node contains the execution logic. It reads the plan and past\_steps from the state, identifies the next step to be executed, and invokes a tool-equipped agent to perform that action. The result of the execution is then returned as an update to the past\_steps list in the state~\citep{langgraphPlanandExecute}.
\item \textbf{replan\_node}: In advanced implementations, a re-planning node can be added. This node examines the history of past\_steps to determine if the execution is proceeding as expected. If a failure occurred or the plan is no longer viable, this node can invoke the planner LLM again to generate a revised plan~\citep{langgraphPlanandExecute}.
\end{itemize}

\paragraph{The Edges} Edges define the control flow, connecting the nodes. LangGraph uses both standard edges and conditional edges. The flow starts with an edge from the START node to the planner\_node. The planner then connects to the executor\_node. The most critical part of the P-t-E graph is the \textbf{conditional edge} that follows the executor. This edge contains logic to inspect the state and decide where to go next. For example, it checks if there are more steps left in the plan. If yes, it routes the flow back to the executor\_node to continue the loop. If all steps are complete, it routes the flow to the END node, terminating the workflow~\citep{langchain-aiLangchainaiLanggraphDiscussions}.

\subsection{Code Reference: A Secure LangGraph Plan-and-Execute Agent}

A complete, runnable example of a secure P-t-E agent built with LangGraph is available in Appendix A.1. This implementation goes beyond basic examples by incorporating the Principle of Least Privilege directly into the executor node.

\paragraph{Security Enhancement} The planner is prompted to produce a plan where each step includes not just a description but also the single specific tool required for that step. The executor node then dynamically creates a new, temporary agent for each step, equipping it only with the one tool specified in the plan. This prevents a tool that is valid for Step 3 from being accessible during the execution of Step 1, providing granular security.

The architecture of LangGraph does more than just implement the P-t-E pattern; it elevates it from a rigid, linear process into a flexible, potentially cyclic graph. This is a fundamental shift that enables the creation of more resilient and intelligent agents. The legacy PlanAndExecute class was largely a ``fire-and-forget'' mechanism - the plan was made and then followed inflexibly~\citep{langchainPlanandExecuteAgents2024}. LangGraph, through its introduction of nodes, state, and particularly conditional edges, makes cycles a natural part of the architecture~\citep{langgraphPlanandExecute}. The most powerful of these cycles is the Execute $\rightarrow$ Re-plan $\rightarrow$ Execute loop. By adding a replan\_node, the agent can inspect the results of its actions (stored in the past\_steps list within the state) and intelligently decide if the original plan remains viable. If a tool fails or returns an unexpected result, the replanner can generate a new, more robust plan, and the conditional edge can route the control flow back to the executor with this revised strategy~\citep{langgraphPlanandExecute}. This transforms P-t-E from a brittle script into an adaptive, self-correcting system, representing a significant leap in agentic robustness and a primary motivation for adopting LangGraph for production systems.

\section{Framework Implementation Guide: CrewAI}

CrewAI is a multi-agent framework designed to facilitate collaboration between autonomous AI agents. Its high-level abstractions - Agents, Tasks, and a Crew - provide a structured and intuitive way to build complex workflows. The framework's design, particularly its hierarchical process, naturally maps to the Plan-then-Execute pattern, with a strong and unique emphasis on granular security controls through task-level tool scoping.

\subsection{The Hierarchical Process: Manager as Planner, Workers as Executors}

CrewAI's architecture is built upon three core components: the Agent, which is the actor performing the work; the Task, which defines a specific unit of work to be done; and the Crew, which orchestrates the agents and tasks to achieve an overall objective~\citep{crewaiHierarchicalProcess}. To implement a P-t-E workflow, one must use the Process.hierarchical setting when instantiating the Crew~\citep{crewaiHierarchicalProcess}.

This setting fundamentally changes the crew's operational model and requires the designation of a manager, which maps directly to the P-t-E components:

\paragraph{Planner $\rightarrow$ Manager Agent} When process is set to hierarchical, the Crew must be assigned a manager\_agent. This agent's designated role is to act as the Planner. It receives the overall goal, analyzes it, breaks it down into a series of discrete tasks, and then delegates each task to the most appropriate worker agent in the crew~\citep{crewaiHierarchicalProcess}. For the manager agent to be able to delegate, its allow\_delegation parameter must be set to True. The manager's system prompt and goal should be crafted to guide it in this strategic planning and coordination role.

\paragraph{Executors $\rightarrow$ Worker Agents} The other agents defined in the agents list of the crew function as the Executors. Each worker is typically designed as a specialist with a specific role, backstory, and a set of tools tailored to its function (e.g., a ``Researcher'' with search tools, a ``Writer'' with file I/O tools)~\citep{crewaiHierarchicalProcess}. They do not create the high-level plan; they receive specific tasks delegated to them by the Manager Agent and execute them.

This manager-worker dynamic provides a clean and high-level abstraction for the P-t-E pattern, allowing developers to focus on defining the roles and capabilities of their agents rather than the low-level mechanics of the control flow.

\subsection{Secure Tool Scoping: Assigning Capabilities to Agents vs. Tasks}

A standout feature of CrewAI, and one that is critical for security, is its mechanism for scoping tool access. The framework makes a crucial distinction between assigning tools at the Agent level versus the Task level, providing a powerful method for enforcing the Principle of Least Privilege~\citep{russoToolBestPractice2025}.

\paragraph{Agent-Level Tools (agent.tools)} This parameter defines the complete set of tools that an agent could use. It represents the agent's full potential capabilities - its entire toolbox~\citep{russoToolBestPractice2025}. For example, a DataAnalyst agent might have tools for reading databases, performing calculations, and writing files.

\paragraph{Task-Level Tools (task.tools)} This parameter defines a specific, limited subset of tools that are permitted for the execution of that particular task. This is the key to granular security control~\citep{russoToolBestPractice2025}.

The critical rule of precedence is that Task.tools overrides Agent.tools. If Task.tools is explicitly defined for a given task, the agent assigned to that task can only use the tools in that list, regardless of what other tools it may have in its own tools list. If the Task.tools list is left undefined, the task will inherit the full set of tools from the assigned agent~\citep{russoToolBestPractice2025}.

This mechanism allows for the implementation of a robust security posture. The best practice is to define a worker agent with its potential capabilities, but then for each task delegated to it, to strictly constrain the available tools to the absolute minimum required. For instance, a FinancialAgent might have both a market\_data\_lookup\_tool and a trade\_execution\_tool. For a ``Market Research'' task, its tools can be scoped to only the market\_data\_lookup\_tool. For a subsequent ``Execute Trade'' task, it can be given access to only the trade\_execution\_tool. This prevents the agent from being tricked into executing a trade while it is only supposed to be conducting research, effectively mitigating the risk of unauthorized tool use~\citep{russoToolBestPractice2025}.

\subsection{Secure Tool Scoping: Hybrid approaches}

CrewAI already enables fine-grained tool restriction on a per-task basis, supporting the enforcement of least privilege dynamically during execution. However, this can be enhanced further through the integration with existing RBAC ecosystems.

In an RBAC-integrated setup, each agent could be tagged with a role at plan time, assigned by the Planner, and each role maps to a static permission profile. For example, the role WebSearcher may include access to web search, summarization, and citation tools, while FinanceWriter may allow read-only access to internal datasets and spreadsheet generation, but no write access to other types of APIs.

This dual enforcement mechanism offers defense-in-depth:

\begin{itemize}
\item Roles act as coarse-grained, pre-approved permission containers, defining what an agent is ever allowed to do.
\item Tasks scope tools within those roles to the minimum necessary for the current execution step.
\end{itemize}

RBAC has proven itself effective in real-world access control at scale, from Kubernetes clusters and AWS IAM policies to internal enterprise tools and SaaS platforms. Its adoption in LLM agent architectures would bring similar benefits: auditability (each action can be logged against the role it was performed under), composability (roles are reusable across agents), and clarity of trust boundaries (system operators can reason about what classes of actions an agent is even capable of). For example, a Planner could output the following metadata during planning:

\begin{verbatim}
{
  "task": "extract_customer_feedback",
  "agent": "feedbackBot",
  "role": "DataReader",
  "tools": ["read_file", "summarise_text"]
}
\end{verbatim}

Here, the role DataReader would be validated against a predefined permission map. Any attempt by this agent to use write or external communication tools would be rejected at execution time, even if they appear in the plan, because they exceed the privileges of the assigned role.

By embedding RBAC directly into the planning and execution lifecycle, agentic systems gain structural enforcement of least privilege that scales with system complexity and existing organizational policies. 

\subsection{Code Reference: A Secure CrewAI Manager-Worker Crew}

A complete, runnable example of a secure P-t-E implementation in CrewAI is available in Appendix A.2. It demonstrates the manager-worker pattern and explicitly uses task-level tool scoping to enforce the Principle of Least Privilege.

The CrewAI design philosophy introduces a subtle but powerful shift in where security controls are applied. In many agentic frameworks, security policies like tool access are primarily bound to the Agent definition. This can be somewhat static; the agent has a fixed set of capabilities. CrewAI, through its powerful Task abstraction, makes security more dynamic and granular. The Task object becomes the primary locus of control, capable of overriding an agent's default toolset for a specific job~\citep{russoToolBestPractice2025}.

This means the same Agent can be deployed in multiple contexts with different, situationally appropriate security postures. A DataAnalyst agent might be defined with a full suite of tools for reading from and writing to a database. However, when the Manager assigns it a ``Generate Quarterly Summary Report'' Task, its tools can be dynamically restricted to read-only database access. Later, when the same agent is assigned an ``Update Customer Records'' Task, it can be granted the necessary write permissions for that specific job. The Planner (the Manager Agent) is responsible for creating and delegating these well-defined, security-scoped Tasks. Consequently, security in CrewAI is not just about defining safe agents; it is about defining safe units of work. This task-centric security model is more flexible, more dynamic, and aligns more closely with the Principle of Least Privilege, as permissions are granted ``just-in-time'' for a specific task, not ``just-in-case'' for the agent's entire operational lifecycle.

\section{Framework Implementation Guide: AutoGen}

Microsoft's AutoGen ~\citep{autogenGroupChatAutoGen} is a highly flexible framework designed for orchestrating complex conversations and workflows between multiple, specialized agents~\citep{awanAutoGenTutorialBuild}. Unlike frameworks with a built-in, high-level primitive for ``plan-and-execute,'' AutoGen achieves this pattern through its core strength: the sophisticated management of multi-agent dialogues. Implementing P-t-E in AutoGen ~\citep{autogenGroupChatAutoGen} involves architecting a conversation flow that enforces the desired separation between planning and execution. This section explores how to construct such a workflow and highlights AutoGen's critical, built-in feature for secure code execution.

\subsection{Orchestrating P-t-E with Group Chat and Sequential Conversations}

AutoGen's fundamental paradigm is the ``conversable agent''~\citep{awanAutoGenTutorialBuild}. The framework's power lies not in predefined workflow structures but in the developer's ability to define the rules of interaction between these agents. There are two primary methods for orchestrating a P-t-E pattern.

\paragraph{Method 1: Sequential Chat (initiate\_chats)}
A straightforward way to mimic a P-t-E workflow is by using the initiate\_chats method, which allows one agent to start a series of distinct, sequential conversations with other agents~\citep{autogenConversationPatternsAutoGen}. The output or summary of one conversation can be passed as a ``carryover'' context to the next.

\begin{itemize}
\item \textbf{Chat 1 (Planning):} A UserProxyAgent (representing the user or a higher-level orchestrator) initiates a chat with a PlannerAgent. The PlannerAgent's sole task is to receive the user's objective and generate a structured plan. The result of this chat - the plan itself - is captured and used as the input for the next stage.
\item \textbf{Chat 2 (Execution):} The UserProxyAgent then initiates a new chat with an ExecutorAgent. The plan from the first chat is passed into this conversation as context. The ExecutorAgent is then responsible for carrying out the steps of the plan. This can be implemented as a loop within the agent's logic or, for more complex plans, as a further sequence of chats.
\end{itemize}

\paragraph{Method 2: Group Chat with Custom Speaker Selection (Recommended)}
This is a more powerful, flexible, and robust method for implementing P-t-E. It involves creating a GroupChat containing all the necessary agents (planner, executors, user proxy) and controlling the flow of conversation via a GroupChatManager equipped with a custom speaker\_selection\_method~\citep{autogenGroupChatCustomized}. This custom function acts as the central orchestrator, effectively hard-coding the P-t-E logic into the state machine of the conversation.

\begin{itemize}
\item \textbf{Agent Roles:} The group would contain a Planner agent, one or more specialist Executor agents (e.g., a CodeWriterAgent), and a UserProxyAgent to execute code and represent the user.
\item \textbf{Custom Logic:} The speaker\_selection\_method is a Python function that is called after each turn to decide which agent speaks next. The logic would be defined as follows:
\begin{enumerate}
\item If the chat has just started, select the Planner agent.
\item If the last speaker was the Planner, analyze its message. If it contains a complete plan, select the appropriate Executor agent.
\item If the last speaker was an Executor that produced code, select the UserProxyAgent (configured as the code executor) to run the code.
\item If the UserProxyAgent reports a successful execution, the function can select the next Executor or terminate the chat. If it reports an error, it can select the CodeWriterAgent again to debug the code.
\end{enumerate}
\end{itemize}

This method turns the conversational flow into a deterministic state machine that precisely follows the P-t-E pattern, including potential loops for debugging and refinement.

\subsection{Enforcing Security with Dockerized Code Execution}

A critical security feature of AutoGen ~\citep{autogenGroupChatAutoGen} is its built-in support for sandboxed code execution. In many AutoGen workflows, the UserProxyAgent is configured not just to represent the human user but also to act as the entity that executes code generated by other agents~\citep{tizkovaMicrosoftsAutoGenGuide}. This is managed through the code\_execution\_config parameter.

This parameter is a dictionary that contains several key settings, but the most important for security is use\_docker~\citep{autogenGroupChatAutoGen}.

\textbf{use\_docker: True:} When this setting is enabled, AutoGen will not execute any generated Python code directly on the host machine's operating system. Instead, it will automatically perform the following actions:

\begin{enumerate}
\item Provision a new Docker container using a specified or default Python image.
\item Execute the generated code snippet inside this isolated container.
\item Capture the results (e.g., stdout) from the container.
\item Return the results to the agent conversation.
\item Terminate and discard the container.
\end{enumerate}

This provides a powerful layer of isolation. Even if a planner or writer agent is compromised through a prompt injection attack and generates malicious code (e.g., os.system('rm -rf /')), the code's execution is contained entirely within the ephemeral Docker environment. It cannot access the host file system, network (unless explicitly configured), or other processes. This makes the use\_docker: True setting an essential, non-negotiable best practice for any AutoGen application that involves code generation and execution.

\subsection{Code Reference: A Secure AutoGen. Planner-Executor Group}

A complete, runnable example of a secure P-t-E workflow in AutoGen~\citep{autogenGroupChatAutoGen}. is available in Appendix A.3. It uses the recommended Group Chat with Custom Speaker Selection method and enforces secure, Dockerized code execution.

The conversational architecture of AutoGen necessitates a unique perspective on implementing agentic patterns like P-t-E. It is less about defining a static data flow graph and more about designing a governed dialogue. The security and logic of the system are emergent properties of the rules of this conversation and the intrinsic capabilities of its participants.

An architect using AutoGen must essentially design a secure protocol. Who is allowed to speak to whom, and in what order? This is managed by the GroupChatManager and the speaker selection logic~\citep{autogenGroupChatCustomized}. What are the capabilities of each participant? This is defined in their agent configurations. Crucially, in what environment do they perform their actions? This is controlled by settings like code\_execution\_config~\citep{autogenGroupChatAutoGen}. The P-t-E pattern is thus abstracted into a problem of protocol design. This mental model is particularly powerful for security engineers, who are well-versed in analyzing and designing secure communication protocols. The overall security of the AutoGen system becomes a function of the robustness of its defined conversational rules and the sandboxed environments in which its agents operate.

\section{Advanced Patterns and Strategic Considerations}

While the basic Plan-then-Execute pattern provides a solid foundation for building predictable and secure agents, its simple, linear form has limitations in complex, real-world scenarios. To create truly resilient, efficient, and trustworthy production-grade systems, architects must incorporate more sophisticated patterns that address brittleness, performance bottlenecks, and the inherent fallibility of LLMs. This section explores advanced strategies, including dynamic re-planning, parallel execution via Directed Acyclic Graphs (DAGs), and the critical role of Human-in-the-Loop (HITL) verification.

\subsection{Dynamic Adaptation: Implementing Re-planning Loops}

A static, one-shot plan is inherently brittle. It assumes a perfect world where tools always succeed, data is always in the expected format, and the initial assumptions made by the planner hold true throughout execution. In reality, it rarely is. A tool call might fail due to a network error, an API may return an unexpected response, or the information gathered in an early step may invalidate the logic of later steps~\citep{langchainPlanandExecuteAgents2024}. Without a mechanism to adapt, the agent will fail.

The ability to dynamically re-plan is, therefore, crucial for building resilient agents~\citep{langchainPlanandExecuteAgents2024}. This involves adding a feedback loop to the P-t-E architecture. The implementation typically involves a ``re-planner'' step or node in the workflow. After each execution step, the control flow moves to this re-planner, which is an LLM call that is provided with:

\begin{enumerate}
\item The original high-level objective.
\item The original plan.
\item The history of all previously executed steps and their outcomes.
\end{enumerate}

With this context, the re-planner LLM can assess the current situation and make an intelligent decision: continue with the existing plan, generate a completely new plan to overcome an obstacle, or determine that the task is complete~\citep{langgraphPlanandExecute}. Frameworks like LangGraph, with their native support for cyclic graphs, are exceptionally well-suited for implementing this pattern. A dedicated replan node can be added after the executor node, with conditional edges routing the flow back to the executor (with a potentially new plan) or to the end of the workflow. This transforms the agent from a rigid automaton into a resilient problem-solver.

\subsection{Enhancing Performance}

\subsubsection{Parallel Execution with Directed Acyclic Graphs (DAGs)}

The standard P-t-E model executes plan steps sequentially, one after another. This is a significant performance bottleneck if the plan contains independent tasks that could be performed concurrently~\citep{langgraphPlanandExecute}. For example, a plan to ``Research competitor A's market share and competitor B's latest product launch'' involves two independent web search tasks that do not need to wait for each other.

To overcome this, the planning model can be advanced from generating a simple linear list to producing a Directed Acyclic Graph (DAG). In a DAG-based plan, each task explicitly declares its dependencies on other tasks~\citep{langchainPlanandExecuteAgents2024}. For example, a task to ``Synthesize research findings'' would have dependencies on the two research tasks mentioned above.

This concept has been formalized in research like the LLMCompiler paper~\citep{langchainPlanandExecuteAgents2024}. In this architecture, the Planner streams a DAG of tasks. A separate component, the ``Task Fetching Unit,'' monitors the graph and schedules tasks for execution as soon as their declared dependencies have been met. This allows for maximal parallelism, as multiple independent tasks can be executed concurrently. The paper reports significant speed boosts (up to 3.6x) from this parallel execution, especially for tasks that involve I/O-bound operations like web searches or API calls~\citep{langchainPlanandExecuteAgents2024}. While implementing a full-scale DAG-based execution engine is a complex engineering task, the architectural concept is critical for developers building performance-sensitive agentic systems.

The evolution of the P-t-E pattern from a simple, linear sequence to a robust loop with re-planning, and finally to a high-performance parallel graph mirrors the historical evolution of software build systems in traditional engineering. The initial P-t-E model is analogous to a simple shell script that executes commands in a fixed order. The introduction of re-planning loops is akin to adding error handling and conditional logic to that script, making it more resilient. The final step, moving to a DAG-based planner that manages dependencies and enables parallel execution, is conceptually identical to the shift from simple scripts to sophisticated build tools like make or Bazel. These tools parse a dependency graph (Makefile or BUILD files) to execute compilation and testing tasks in parallel for maximum efficiency. This parallel suggests that as agentic systems mature, they are independently rediscovering and adapting proven principles from decades of software architecture. Architects can leverage this understanding by applying established patterns like dependency management, parallel processing, and fault tolerance to their agent designs, rather than treating them as an entirely novel class of problems.

\subsubsection{Minimizing Context Overhead with GraphQL Integration}

Another promising direction for optimizing data retrieval in multi-agent systems is the integration of GraphQL~\citep{graphqlGraphQLQueryLanguage} as a tool interface. While not yet widely adopted in current multi-agentic frameworks, it offers a highly aligned model for P-t-E architectures: agents can request exactly the fields they need from structured data sources, minimizing context window usage and API latency.

In this design, the Planner could formulate or select GraphQL queries (e.g., through a schema-aware tool wrapper), and the Executor would execute them as discrete steps. This has two clear benefits:

\begin{itemize}
\item \textbf{Reduced token overhead}, since only the explicitly requested fields (e.g., user.name rather than the entire profile) are returned and added to context
\item \textbf{Fewer steps per plan}, as GraphQL can resolve nested or related data in a single call, avoiding the need to chain multiple REST tool invocations.
\end{itemize}

Emerging frameworks such as MCP (and its multi-server architecture) already demonstrate how GraphQL operations can function as callable tools~\citep{hawkinsEveryTokenCounts} This confirms the feasibility of treating GraphQL queries as structured, predictable agents' tools, delivering JSON payloads that are compact, auditable, and easy to constrain via schema validation.

While a full GraphQL agent integration remains a forward-looking recommendation, it represents a clean, cost-efficient path for multi-agent architectures that require consistent access to structured internal data, particularly in systems where token control, latency, and interface trust boundaries are key concerns.

\subsubsection{Graph-Based Conditional Execution Paths}

Rather than waiting for failures to occur before re-engaging the Planner, agentic systems can be made more resilient by embedding alternate execution paths directly into the initial plan structure. One way to improve execution resilience is to embed conditional fallback branches directly into the initial plan, allowing agents to handle known contingencies without restarting from scratch. Instead of outputting a strictly linear execution sequence, the Planner could emit a graph that includes aforementioned conditional edges, encoding alternate paths or recovery actions ahead of time.

This approach builds upon the graph model discussed in the previous section, extending it beyond parallelism into dynamic routing. For example, the Planner could produce logic like: ``If Tool A fails, attempt Tool B,'' or ``If the input data is structured as X, follow path A; else follow path B.'' These embedded decision points would enable the Executor to locally pivot to alternate steps when minor, anticipated failures occur, without re-invoking the Planner or invalidating the entire plan.

As mentioned before, LangGraph already supports this model through conditional edge evaluation (you can see an example in the Appendix), making it feasible to express branching control flow in LLM-driven workflows~\citep{ai/mlLangGraphSimplifiedUnderstanding2025} By leveraging these capabilities, developers can predefine multiple valid routes through the plan graph, allowing agentic systems to recover gracefully from predictable faults while still reserving full re-planning for true edge cases. This pattern saves latency, reduces token overhead, and enhances robustness by turning the plan into a resilient decision graph, rather than a fragile one-shot instruction list.

We recommend that for high-critical or extensive planning tasks, developers consider encoding such fallback logic during planning, either through conditionally linked nodes or alternate execution branches that are activated based on runtime signals. In doing so, the P-t-E paradigm gains a valuable middle ground: not just rigid upfront planning or expensive full re-plans, but intelligent local adaptation built into the plan itself.

\subsection{Calibrating Trust: The Role of Human-in-the-Loop (HITL) Verification}

Deploying LLM agents in real-world applications requires a carefully calibrated level of trust between the human user and the AI system. Research has shown that agents can be a ``double-edged sword''~\citep{hePlanThenExecuteEmpiricalStudy2025}. They are immensely capable, but they can also be ``convincingly wrong'', generating plans, code, or factual statements that appear plausible, confident, and well-structured but are fundamentally incorrect or logically flawed~\citep{hePlanThenExecuteEmpiricalStudy2025}. This can mislead users into placing unwarranted trust in a faulty agent.

Human-in-the-Loop (HITL) verification is the primary mechanism for mitigating this risk and calibrating user trust. However, empirical studies show that the effectiveness of HITL depends heavily on where in the workflow it is applied~\citep{hePlanThenExecuteEmpiricalStudy2025}.

\paragraph{User-Involved Planning} Contrary to intuition, directly involving a user in the initial planning phase does not consistently lead to better plans. In some cases, users can be swayed by the LLM's confident but incorrect suggestions and may even introduce new errors while trying to ``help'' modify the plan~\citep{hePlanThenExecuteEmpiricalStudy2025}. The risk of being misled by a ``convincingly wrong'' plan is high at this stage.

\paragraph{User-Involved Execution} HITL is far more beneficial during the execution phase. When presented with a specific, concrete action and its outcome (e.g., ``The agent is about to send this email. Approve?''), users are effective at catching specific errors, correcting faulty tool outputs, and preventing undesirable actions. This is particularly true for tasks with higher complexity or risk~\citep{hePlanThenExecuteEmpiricalStudy2025}.

Based on these findings, the recommended architectural pattern for many critical systems is \textbf{Automatic Planning + User-Involved Execution}~\citep{hePlanThenExecuteEmpiricalStudy2025}. This leverages the LLM's strength in generating a comprehensive initial strategy while using the human's superior judgment to supervise the tactical execution of each step. As discussed previously, this can be further strengthened for the highest-risk scenarios by evolving the pattern to \textbf{Plan-Validate-Execute}, where an expert human must approve the automatically generated plan \textit{before} any execution begins, providing a crucial safeguard against acting on a fundamentally flawed strategy.

\subsubsection{HITL vs Automated Verification}

While human review provides a strong safeguard against flawed reasoning, it does not always scale, particularly in automated or low-risk workflows. To address this, we could augment the architecture with an automated Verifier agent (see Section 1), a distinct module responsible for validating the Planner's output before execution begins. This pattern could extend P-t-E into a Plan-Validate-Execute loop, where validation occurs independently of planning and execution.

As mentioned, the Verifier can be instantiated as another LLM (with a different prompt or model), a rule-based engine, or a symbolic checker. Its role is to inspect the plan for logical coherence, policy compliance, and alignment with system constraints. This introduces a layer of redundancy and cross-checking that helps mitigate issues like overconfidence or hallucination in the Planner's output.

This pattern aligns with recent architectural advances such as the use of dedicated \textbf{process verifiers}, agents that review intermediate reasoning steps to ensure progress toward a goal, rather than trusting the planner's output wholesale. For example, this paper~\citep{setlurRewardingProgressScaling2024} introduces a similar idea, demonstrating how separate verifier agents can guide and validate structured reasoning processes.

Incorporating a Verifier improves system safety without requiring human oversight for every single task. In enterprise or compliance-sensitive contexts, it also creates an internal audit checkpoint, making failures easier to diagnose and trust boundaries easier to enforce.

\subsection{Architectural Weaknesses: The Cost of Upfront Planning}

While the Plan-then-Execute pattern offers significant advantages in predictability, security, and reasoning quality for complex tasks, it is not without its architectural weaknesses, particularly concerning initial latency and cost-efficiency. These trade-offs stem directly from its defining feature of comprehensive upfront planning.

\subsubsection{Upfront Latency and ``Time-to-First-Action''}
The most immediate drawback of the P-t-E pattern is its initial response time. Before the agent can perform its first action, it must make a call to a powerful planner LLM and wait for the entire multi-step plan to be generated. This planning phase can introduce significant latency, making the pattern feel slower to the end-user compared to more reactive approaches~\citep{pattenLLMSecuritySafe2025}. A ReAct agent, in contrast, performs a much quicker initial reasoning step and can execute its first tool call almost immediately. For applications where a fast ``time-to-first-action'' is critical, the P-t-E pattern's initial planning overhead is a distinct disadvantage.

\subsubsection{High Token Consumption in the Planning Phase}
The comprehensive nature of the planning phase often leads to higher initial token consumption. The prompt sent to the planner must contain the full user objective, detailed instructions, and tool definitions. The resulting plan can be lengthy and verbose. For complex tasks, this single planning call can consume a large number of tokens, sometimes between 3,000 and 4,500, which can be more than a ReAct agent might use for an entire simple task~\citep{pattenLLMSecuritySafe2025}. This front-loaded cost can be a significant concern for cost-sensitive applications, especially if many of the tasks are simple enough that they do not warrant such an extensive planning investment.

\paragraph{Hierarchical Planning through Sub-Planners} 
While the conventional P-t-E architecture introduces a single Planner responsible for generating the full execution roadmap, in more complex or cost-sensitive workflows, this responsibility can be distributed across multiple planning layers. An architectural extension involves introducing sub-planners - smaller, purpose-specific planning agents that handle discrete segments of the overall task. In this hierarchical model, the primary \textbf{Planner first generates a high-level blueprint by decomposing the user's objective into broad milestones or phases}. These milestones are then delegated to sub-planners, which operate independently or in parallel to develop detailed plans for their assigned segments.
This design offers several advantages. First, it reduces upfront token costs (as sub-planners operate with narrower context windows and more focused objectives), avoiding a monolithic, memory-heavy planning prompt. Second, it enhances modularity and reusability; sub-planners can be specialized for domain-specific planning (e.g., data preprocessing, API integration, or report generation), allowing for targeted optimization or fine-tuning. Finally, it introduces a natural failure containment boundary: if one sub-plan is flawed or requires re-planning, only that segment needs regeneration, preserving the rest of the execution logic and improving overall system resilience. This hierarchical approach could extend the P-t-E pattern from a linear, centralized planning model to a distributed, scalable planning ecosystem more suitable for real-world, multi-agent applications.

\subsubsection{Risk of Wasted Effort and Sequential Bottlenecks}
The upfront investment in planning carries the risk of being entirely wasted. If the initial plan is flawed or an early step fails in an unrecoverable way (and the system lacks a re-planning loop), the time and tokens spent on generating the full plan are lost. Furthermore, in its basic form, the P-t-E pattern executes its plan sequentially~\citep{langgraphPlanandExecute}. This creates an unnecessary performance bottleneck if the plan contains independent tasks that could have been executed in parallel. While advanced implementations using DAGs can mitigate this, the standard linear execution model remains a potential weakness~\citep{langgraphPlanandExecute}.  

Ultimately, the P-t-E pattern's cost and latency profile represent a strategic trade-off. It front-loads the computational expense with the expectation that this will lead to a more efficient and successful overall execution for complex, multi-step workflows~\citep{singhPlanExecuteAI2024}. However, for simpler tasks or scenarios where initial responsiveness is paramount, this upfront cost can become a significant architectural liability, making a more iterative pattern like ReAct a more suitable choice~\citep{pattenLLMSecuritySafe2025}.

\section{Strategic Recommendations for Implementation}

The successful deployment of secure and reliable LLM agents requires a combination of architectural theory, security-first principles, and practical framework knowledge. The choice of a reasoning pattern and implementation framework is a critical decision with long-term consequences for a project's flexibility, scalability, and security posture. This final section provides a consolidated framework selection matrix and a strategic architectural blueprint to guide architects, developers, and security engineers in building production-grade Plan-then-Execute agents.

\subsection{Framework Selection Matrix: Matching Tools to Use Cases}

The choice between LangGraph, CrewAI, and AutoGen. depends on the specific priorities of the project. Each framework offers a different set of abstractions and trade-offs for implementing the P-t-E pattern. A comparative analysis to aid in this strategic decision is provided in Appendix B.3.

\subsection{Architectural Blueprint for Secure, Production-Grade Agents}

Regardless of the chosen framework, a set of core architectural principles should guide the development of any secure, production-grade agentic system. This report's key findings are summarized in the following recommendations, forming an actionable blueprint.

\begin{enumerate}
\item \textbf{Default to Plan-then-Execute:} For any task that is non-trivial or involves more than one or two tool calls, the Plan-then-Execute pattern should be the default high-level architecture. Its inherent predictability, control, and resistance to control-flow hijacking make it a more suitable foundation for reliable systems than purely reactive patterns.
\item \textbf{Embrace Defense-in-Depth:} No single pattern is a silver bullet~\citep{phdSandboxedMindPrincipled2025}. A robust security posture is achieved by layering multiple controls. The P-t-E pattern must be combined with a comprehensive set of complementary security measures, including:
\begin{itemize}
\item \textbf{Strict Input Sanitization and Output} Filtering to protect the data plane.
\item \textbf{The Principle of Least Privilege} applied rigorously to tool access, ideally at the task or step level.
\item \textbf{Strongly Sandboxed Execution} for any code generation capabilities, with Docker being the industry standard.
\end{itemize}
\item \textbf{Design for Resilience with Re-planning:} A static plan is a brittle plan. Production systems must be able to handle unexpected failures and adapt. Architect the workflow to include a re-planning loop that allows the agent to assess failures in the context of its overall goal and formulate a new strategy.
\item \textbf{Implement Risk-Appropriate Human Oversight:} Calibrate the level of human involvement based on the risk and criticality of the agent's tasks.
\begin{itemize}
\item For low-risk tasks, fully autonomous operation may be acceptable.
\item For medium-risk or irreversible actions, implement \textbf{User-Involved Execution}, requiring human approval for specific steps.
\item For high-stakes, mission-critical systems, adopt the \textbf{Plan-Validate-Execute} pattern, where an expert human must validate the agent's entire plan before any execution is permitted.
\end{itemize}
\item \textbf{Ensure Comprehensive Auditability:} The ability to understand why an agent took a certain action is critical for debugging, compliance, and trust. Implement robust, structured logging for all significant events, including the initial user request, the generated plan, each tool call with its inputs and outputs, any re-planning decisions, and the final response. Observability platforms like LangSmith are invaluable for tracing these complex execution paths~\citep{LangchainaiLanggraph2025}.
\end{enumerate}
Ultimately, the challenge of building secure LLM agents is a systems architecture problem, not merely a model-tuning problem. By moving beyond attempts to make the LLM itself infallible and instead focusing on building a resilient and constrained architectural framework around it, we can harness the power of these models. By applying proven engineering principles of security, modularity, fault tolerance, and resilience, developers and architects can construct powerful, effective, and trustworthy agentic systems capable of tackling real-world challenges.

\nocite{willisonDesignPatternsSecuring,singhLLMBasedAgentsBenefits,hinchyHypeHowSecurity,zhengTracingAutoGenPrompt,pattenLLMSecuritySafe2025,langchain-aiLangchainaiLanggraphDiscussions,LangchainaiLanggraph2025,teetrackerCrewAIHierarchicalManager2025,awanAutoGenTutorialBuild,t.ph.d.CrewAIsTaskTool2024,crewaiCustomManagerAgent,crewaiCustomizeAgents}
\nocite{dasHowBuildLLM}
\nocite{autogenConversationPatternsAutoGen}
\nocite{iraniTutorialMultiAgentInteractions2024}
\nocite{winlandWhatCrewAIIBM2024,sugarforeverAutoGenTutorialsAutogen_rag_agentipynbMain,sinanuOreillyaiagentsNotebooksLangGraph_Plan_Executeipynb,sahotaPlanandExecuteAgentsLangchain2023,padmanabhanPlanandExecuteLangChainHandling2025,ngAgenticDesignPatterns2024,manikaHowBuildAI,autogenGroupChatCoder,crewaiTasks,deepchartsBuildPlanExecute2025,fabkostaHowDoesLangChain2023a,langchain-aiLangchainCookbookPlan_and_execute_agentipynb}


\bibliography{tmlr}
\bibliographystyle{tmlr}

\appendix

\section{Code Examples}

\subsection{LangGraph Plan-and-Execute Agent}

\begin{python}
import os
import operator
from typing import TypedDict, Annotated, List, Tuple, Union

from langchain_core.pydantic_v1 import BaseModel, Field
from langchain_core.prompts import ChatPromptTemplate
from langchain_openai import ChatOpenAI
from langgraph.graph import StateGraph, END
from langgraph.prebuilt import create_react_agent
from langchain_core.tools import tool

# --- 1. Define Tools ---
# Define a few distinct tools to demonstrate security scoping.

@tool
def web_search(query: str) -> str:
    """Performs a web search for the given query."""
    # In a real implementation, this would call a search API.
    print(f"--- Executing Web Search for: {query} ---")
    if "weather" in query.lower():
        return "The weather in San Francisco is 65°F and sunny."
    elif "capital of france" in query.lower():
        return "The capital of France is Paris."
    return "No relevant information found."

@tool
def write_file(filename: str, content: str) -> str:
    """Writes the given content to a file."""
    print(f"--- Writing to file: {filename} ---")
    with open(filename, "w") as f:
        f.write(content)
    return f"Successfully wrote to {filename}."

@tool
def calculate(expression: str) -> str:
    """Calculates the result of a mathematical expression."""
    print(f"--- Calculating: {expression} ---")
    try:
        result = eval(expression, {"__builtins__": {}}, {})
        return f"The result is {result}."
    except Exception as e:
        return f"Error during calculation: {e}"
# Master list of all available tools
all_tools = [web_search, write_file, calculate]
tool_map = {t.name: t for t in all_tools}

# --- 2. Define Graph State ---
# The state object that will be passed between nodes.

class PlanExecuteState(TypedDict):
    input: str
    plan: List[dict]  # Each dict will have 'task' and 'tool_name'
    past_steps: Annotated], operator.add]
    response: str

# --- 3. Define Planner and Executor Logic ---

# Pydantic models for structured output from the planner
class PlanStep(BaseModel):
    task: str = Field(description="The detailed description of the sub-task to perform.")
    tool_name: str = Field(description=f"The single, most appropriate tool to use for this task from the available tools: {[t.name for t in all_tools]}")

class Plan(BaseModel):
    """A comprehensive, step-by-step plan to achieve the user's objective."""
    steps: List = Field(description="The sequence of steps to execute.")

# Planner Node
def planner_node(state: PlanExecuteState):
    """
    Generates a plan with a specific tool assigned to each step.
    """
    print("--- Planning... ---")
    prompt = ChatPromptTemplate.from_template(
        """For the given objective, create a detailed, step-by-step plan.
        For each step, specify the single tool from the available list that is best suited to perform the task.
        Objective: {input}
        Available Tools: {tools}"""
    )
    planner_llm = ChatOpenAI(model="gpt-4o", temperature=0)
    structured_planner = planner_llm.with_structured_output(Plan)
    
    plan_result = structured_planner.invoke({
        "input": state["input"],
        "tools": ", ".join([t.name for t in all_tools])
    })
    
    # Convert Pydantic models to simple dicts for state compatibility
    plan_as_dicts = [{"task": step.task, "tool_name": step.tool_name} for step in plan_result.steps]    
    return {"plan": plan_as_dicts}

# Executor Node
def executor_node(state: PlanExecuteState):
    """
    Executes the next step in the plan using ONLY the specified tool for that step.
    """
    plan = state["plan"]
    past_steps = state["past_steps"]
    
    if len(past_steps) >= len(plan):
        return {"response": "Finished all planned steps."}

    current_step_dict = plan[len(past_steps)]
    task_description = current_step_dict["task"]
    tool_name = current_step_dict["tool_name"]
    
    print(f"--- Executing Step {len(past_steps) + 1}: {task_description} ---")
    print(f"--- Security Constraint: Using ONLY tool '{tool_name}' ---")

    # Security Enhancement: Provide only the single, specified tool to the executor agent.
    if tool_name not in tool_map:
        result = f"Error: Tool '{tool_name}' not found. Available tools: {list(tool_map.keys())}"
    else:
        scoped_tool = tool_map[tool_name]
        
        # Create a temporary, single-tool agent for this step
        executor_agent_llm = ChatOpenAI(model="gpt-4o-mini", temperature=0)
        agent_executor = create_react_agent(executor_agent_llm, [scoped_tool])
        
        # Create a rich prompt for the executor agent
        executor_prompt = f"""You are a task executor. Your goal is to complete the following task.
        Task: '{task_description}'
        
        To do this, you have access to ONLY ONE tool: '{scoped_tool.name}'.
        
        Previous steps have been completed and their results are available if needed, but you should focus on the current task.
        History of past steps: {past_steps}
        
        Execute the task using the provided tool.
        """
        
        result = agent_executor.invoke({"messages": [("user", executor_prompt)]})
        result = result.get("messages", [{}])[-1].content

    return {"past_steps": [(current_step_dict, result)]}

# Conditional Edge Logic
def should_continue(state: PlanExecuteState):
    """
    Determines whether to continue executing the plan or to end.
    """
    if len(state["past_steps"]) >= len(state["plan"]):
        return END
    return "executor"

# --- 4. Construct the Graph ---

workflow = StateGraph(PlanExecuteState)

workflow.add_node("planner", planner_node)
workflow.add_node("executor", executor_node)

workflow.set_entry_point("planner")

workflow.add_edge("planner", "executor")
workflow.add_conditional_edges(
    "executor",
    should_continue,
)

app = workflow.compile()

# --- 5. Run the Agent ---
if __name__ == "__main__":
    # Set your OpenAI API key
    # os.environ["OPENAI_API_KEY"] = "your_api_key_here"
    
    objective = "What is the weather in San Francisco, and what is 100 * 5? Write the final answers to a file named 'results.txt'."
    config = {"recursion_limit": 50}
    
    final_state = app.invoke({"input": objective}, config=config)
    
    print("\n--- Final Result ---")
    print(f"Final Response: {final_state.get('response')}")
    print("Execution History:")
    for step, result in final_state['past_steps']:
        print(f"  - Step: {step['task']} (Tool: {step['tool_name']}) -> Result: {result}")

\end{python}

\subsection{ CrewAI Manager-Worker Crew}

\begin{python}
import os
from crewai import Agent, Task, Crew, Process
from crewai_tools import SerperDevTool, FileWriterTool

# --- 1. Setup Environment ---
# Set API keys for required services.
# os.environ["OPENAI_API_KEY"] = "YOUR_OPENAI_API_KEY"
# os.environ = "YOUR_SERPER_API_KEY"

# --- 2. Define Tools ---
# Define the tools that will be available.
search_tool = SerperDevTool()
file_writer_tool = FileWriterTool()

# --- 3. Define Worker Agents (Executors) ---
# These agents are specialists doing the actual work.

# Researcher Agent
researcher = Agent(
    role="Senior Technology Researcher",
    goal="Discover cutting-edge developments in AI and machine learning.",
    backstory="""You are a renowned researcher at a top-tier technology lab.
    Your expertise lies in identifying emerging trends and explaining their significance.
    You have a knack for finding the most relevant and impactful information.""",
    verbose=True,
    # This agent CANNOT delegate.
    allow_delegation=False,
    # This defines the agent's full potential toolkit.
    tools=[search_tool] 
)

# Writer Agent
writer = Agent(
    role="Technical Content Strategist",
    goal="Craft compelling and clear content from technical research findings.",
    backstory="""You are a professional technical writer known for your ability
    to distill complex topics into accessible and engaging articles.
    You are skilled at structuring information logically and writing to a file.""",
    verbose=True,
    allow_delegation=False,
    # This agent's potential toolkit includes the file writer.
    tools=[file_writer_tool]
)

# --- 4. Define Manager Agent (Planner) ---
# This agent's role is to plan and delegate tasks to the workers.
manager = Agent(
    role="Research Project Manager",
    goal="Efficiently manage the research and writing process to produce a high-quality report.",
    backstory="""You are an experienced project manager, skilled in overseeing complex
    research projects. Your role is to break down the main goal into actionable tasks
    and delegate them to the appropriate team members, ensuring a smooth workflow
    and a final output that meets all requirements.""",
    verbose=True,
    # The manager MUST be able to delegate.
    allow_delegation=True
)

# --- 5. Define Tasks with Scoped Tools ---
# Define the units of work. Crucially, we scope the tools at the task level for security.

# Task 1: Research
# This task is assigned to the researcher agent.
research_task = Task(
    description="""Conduct a thorough search for the latest advancements in Large Language Model (LLM) agents.
    Focus on the 'Plan-then-Execute' architectural pattern. Identify at least 3 key benefits and 3 key security considerations.""",
    expected_output="A detailed report summarizing the findings, including benefits and security considerations.",
    agent=researcher,
    # SECURITY: Even though the researcher agent is defined with the search_tool,
    # we explicitly scope this task to ONLY use the search_tool.
    # If file_writer_tool were in the agent's list, it would be inaccessible for this task.
    tools=[search_tool]
)

# Task 2: Write
# This task is assigned to the writer agent.
write_task = Task(
    description="""Using the research report provided, write a concise summary of the findings.
    The summary should be well-structured, clear, and saved to a file named 'llm_agents_report.md'.""",
    expected_output="The final summary content written to the file 'llm_agents_report.md'.",
    agent=writer,
    # SECURITY: This task is strictly limited to using the file_writer_tool.
    # The writer agent cannot perform a web search during this task, even if it had the tool.
    tools=[file_writer_tool],
    # Specify the output file for this task.
    output_file='llm_agents_report.md'
)

# --- 6. Instantiate the Crew with Hierarchical Process ---
# The crew is configured with the manager and the hierarchical process.
crew = Crew(
    agents=[researcher, writer], # The worker agents.
    tasks=[research_task, write_task], # The tasks to be completed.
    manager_agent=manager, # The designated planner.
    process=Process.hierarchical # Enables the manager-led P-t-E workflow.
)

# --- 7. Kickoff the Crew's Work ---
if __name__ == "__main__":
    result = crew.kickoff()
    print("\n\n########################")
    print("## Crew Execution Result:")
    print("########################\n")
    print(result)

\end{python}

\subsection{AutoGen Planner-Executor Group}

\begin{python}
import os
import autogen.

# --- 1. Setup Environment and LLM Config ---
# Ensure you have a Docker daemon running.
# Set your OpenAI API key, for example:
# os.environ["OPENAI_API_KEY"] = "your_api_key_here"

config_list = autogen.config_list_from_json(
    "OAI_CONFIG_LIST",
    filter_dict={"model": ["gpt-4o"]},
)

llm_config = {"config_list": config_list, "timeout": 120}

# --- 2. Define Agents ---

# Planner Agent: Creates the plan.
planner = autogen.AssistantAgent(
    name="Planner",
    system_message="""You are a meticulous planner. Your goal is to create a step-by-step Python script to solve the user's request.
    Do not write the full code yourself. Instead, provide a clear, numbered plan of what the code should do.
    For example:
    1. Import necessary libraries (e.g., pandas, matplotlib).
    2. Load the data from the given source.
    3. Perform the required analysis or visualization.
    4. Print or save the final result.
    Conclude your plan with the phrase 'Plan complete.' for the next agent to take over.""",
    llm_config=llm_config,
)

# Coder Agent: Writes Python code based on the plan.
coder = autogen.AssistantAgent(
    name="Coder",
    system_message="""You are a Python programmer. You will be given a plan.
    Your task is to write a single, complete Python script to execute that plan.
    Ensure the script is self-contained and ready for execution.
    Do not ask for confirmation. Write the code directly in a Python code block.
    If you need to fix code based on an error, provide the full, corrected script.""",
    llm_config=llm_config,
)

# Executor Agent: Executes the code in a secure environment.
# SECURITY: This agent is configured to execute code within a Docker container.
executor = AutoGen..UserProxyAgent(
    name="Executor",
    human_input_mode="NEVER",
    code_execution_config={
        "work_dir": "autogen_coding",
        "use_docker": True,  # This is the critical security setting.
    },
)

# --- 3. Define Custom Speaker Selection Logic for P-t-E ---
def custom_speaker_selection_func(last_speaker: AutoGen.Agent, groupchat: autogen.GroupChat):
    """A custom function to enforce the Plan -> Code -> Execute workflow."""
    messages = groupchat.messages
    
    # If the chat just started, the Planner begins.
    if len(messages) <= 1:
        return planner
    
    # If the Planner just spoke and finished the plan, the Coder takes over.
    if last_speaker.name == "Planner":
        if "Plan complete." in messages[-1]['content']:
            return coder
        else: # Planner needs to continue planning
            return planner

    # If the Coder just provided a script, the Executor runs it.
    elif last_speaker.name == "Coder":
        return executor
        
    # If the Executor just ran the code, check for errors.
    elif last_speaker.name == "Executor":
        # If there was an error, the Coder needs to fix it.
        if "exitcode: 1" in messages[-1]['content'] or "Error:" in messages[-1]['content']:
            return coder
        else:
            # If successful, the task is done.
            return None # Terminate the chat
            
    else:
        return None # Terminate by default

# --- 4. Create the Group Chat and Manager ---
groupchat = autogen.GroupChat(
    agents=[planner, coder, executor],
    messages=,
    max_round=10,
    speaker_selection_method=custom_speaker_selection_func
)

manager = autogen.GroupChatManager(groupchat=groupchat, llm_config=llm_config)

# --- 5. Initiate the Chat ---
if __name__ == "__main__":
    objective = "Plot a bar chart of the populations of the 5 most populous countries in the world and save it to a file named 'population_chart.png'."
    
    # The executor agent will initiate the chat, but the custom selection function
    # will immediately pass control to the Planner.
    executor.initiate_chat(
        manager,
        message=objective,
    )

\end{python}

\section{Tables}

\subsection{Table 1: ReAct vs. Plan-then-Execute - A Comparative Analysis}

\begin{table}[h!]
\centering
\begin{tabular}{p{2cm} p{4.5cm} p{4.5cm} p{5cm}}
\hline
\textbf{Metric} & \textbf{ReAct Pattern} & \textbf{Plan-then-Execute Pattern} & \textbf{Best For} \\
\hline
Core Loop & Iterative: Thought $\rightarrow$ Action $\rightarrow$ Observation & Sequential: Plan $\rightarrow$ [Execute $\rightarrow$ Observe]* & P-t-E is better for structured, predictable workflows. \\
Task Complexity & Low to Medium. Excels at simple, direct tasks. & High. Excels at complex, multi-step, dependent tasks. & P-t-E is superior for orchestrating long chains of actions. \\
LLM Calls & One LLM call per action. & One LLM call per plan (or re-plan). & P-t-E is more efficient for tasks with many steps. \\
Typical Cost & Lower for short tasks; scales linearly and can become high. & Higher upfront cost for planning; scales sub-linearly. & P-t-E offers better cost control for complex operations. \\
Adaptability & High. Adapts its strategy after every single step. & Low (without re-planning). The plan is fixed upfront. & ReAct is better for highly dynamic or unpredictable environments. \\
Predictability & Low. The agent's path can be emergent and hard to predict. & High. The entire sequence of actions is known in advance. & P-t-E is essential for systems requiring auditable and reliable behavior. \\
Security (Indirect Injection) & More vulnerable. Malicious tool output can hijack the next thought. & More resilient. Control flow is locked before tool execution. & P-t-E provides a strong architectural defense against prompt injection. \\
Error Recovery & Prone to getting stuck in loops without sophisticated handling. & More robust, especially when combined with a re-planning loop. & P-t-E with re-planning offers superior resilience. \\
\hline
\end{tabular}
\caption{Comparison of ReAct vs Plan-then-Execute patterns.}
\label{tab:react_vs_pte}
\end{table}

\subsection{Table 2: Security Threat Mitigation with Plan-then-Execute and Complementary Patterns}

\begin{table}[h!]
\centering
\begin{tabular}{p{2cm} p{4cm} p{4.5cm} p{4.5cm}}
\hline
\textbf{Threat Vector} & \textbf{Description} & \textbf{Primary Mitigation (via P-t-E)} & \textbf{Essential Complementary Controls} \\
\hline
Indirect Prompt Injection & Malicious instructions in tool outputs hijack agent's reasoning loop. & P-t-E Control-Flow Integrity: Plan is fixed before ingesting untrusted tool data, preventing logic hijacking. & Input Sanitization, Output Filtering, Dual LLM Pattern. \\
Unauthorized Tool Use & Agent is tricked into using a tool that is inappropriate for the current task. & N/A (P-t-E alone doesn't scope tools). & Principle of Least Privilege: Tools are scoped to the specific task or step, not the agent globally. \\
Privilege Escalation & Agent is coerced into performing a high-privilege, irreversible action. & P-t-E Plan Review: The plan can be audited before execution to spot risky steps. & Human-in-the-Loop (HITL) Approval: Mandatory human sign-off for critical actions. \\
Malicious Code Execution & Agent generates and runs harmful code on the host system. & P-t-E Plan Review: The plan to generate code can be audited before execution. & Sandboxed Execution: All code execution must occur in an isolated environment, like a Docker container. \\
Data Exfiltration & Agent is tricked into sending sensitive internal data to an external entity. & P-t-E Plan Review: The plan can be audited for steps that involve outbound communication. & Output Filtering, Network Egress Controls, Data Loss Prevention (DLP) systems. \\
\hline
\end{tabular}
\caption{Threat vectors, mitigations via Plan-then-Execute, and complementary controls.}
\label{tab:pte_threats}
\end{table}

\subsection{Table 3: Framework Implementation Comparison for Plan-then-Execute}

\begin{table}[h!]
\centering
\begin{tabular}{p{2.5cm} p{4.5cm} p{4.5cm} p{4.5cm}}
\hline
\textbf{Dimension} & \textbf{LangGraph} & \textbf{CrewAI} & \textbf{AutoGen} \\
\hline
Core Abstraction & State Graph: Low-level, explicit control over state, nodes, and edges. & Crew: High-level abstraction of Agents, Tasks, and a managed Process. & Conversation: Flexible orchestration of dialogues between conversable agents. \\
Planning Mechanism & Custom Planner Node: Developer implements a function that calls an LLM to generate the plan. & Hierarchical Manager Agent: A designated agent with \texttt{allow\_delegation=True} acts as the planner. & Planner Agent in a Governed Chat: A specialized agent whose role is to plan within a custom conversational flow. \\
Execution Control & Executor Node: A function that executes one step, often using a temporary ReAct agent. & Worker Agents executing Tasks: Specialized agents are delegated specific, pre-defined tasks. & Executor Agent in a Governed Chat: An agent (often a \texttt{UserProxyAgent}) executes code or tool calls as a turn in the conversation. \\
Re-planning & Natively Supported: Cyclic graph structure makes re-planning loops a first-class citizen. & Possible via Logic: Requires custom logic within the Manager Agent to re-evaluate and re-delegate tasks. & Possible via Custom Flow: Requires designing the custom speaker selection logic to loop back to the planner on certain conditions. \\
Tool Scoping (Least Privilege) & Programmatic: Must be implemented manually in the executor node by dynamically providing tools. & Declarative: A core feature. \texttt{Task.tools} provides explicit, granular control, overriding \texttt{Agent.tools}. & Via Agent Capabilities: Tools are generally tied to the agent; requires careful agent design or custom logic to scope. \\
Execution Sandboxing & Manual Implementation: Developer is responsible for integrating sandboxing (e.g., Docker). & Manual Implementation: Developer is responsible for ensuring tools that execute code do so safely. & Built-in Support: \texttt{code\_execution\_config} with \texttt{use\_docker=True} provides a first-class, secure sandboxing mechanism. \\
Best For & Maximum Flexibility and Control: Ideal for complex, bespoke workflows requiring fine-grained control over state and logic, and for building resilient, self-correcting agents. & Rapid Development and Clarity: Excellent for quickly building robust multi-agent systems with clear, role-based specializations and strong, declarative security for tools. & Complex Agent-to-Agent Interactions: Best for scenarios where the workflow is best modeled as a sophisticated, multi-party conversation with dynamic turn-taking. \\
\hline
\end{tabular}
\caption{Comparison of LangGraph, CrewAI, and AutoGen. across key dimensions.}
\label{tab:langgraph_crewai_autogen}
\end{table}

\end{document}

%% file: math_commands.tex
\usepackage{amsmath,amsfonts,bm}

\def\eqref#1{equation~\ref{#1}}

\def\1{\bm{1}}

\DeclareMathAlphabet{\mathsfit}{\encodingdefault}{\sfdefault}{m}{sl}
\SetMathAlphabet{\mathsfit}{bold}{\encodingdefault}{\sfdefault}{bx}{n}

